\documentclass[conference]{IEEEtran}
  
\usepackage{epsfig}
\usepackage{times}
\usepackage{float}
\usepackage{afterpage}
\usepackage{amsmath}
\usepackage{mathrsfs}
\usepackage{amstext}
\usepackage{amssymb,bm}
\usepackage{latexsym}
\usepackage{color}
\usepackage{graphicx}
\usepackage{amsmath}
\usepackage{amsthm}
\usepackage{graphicx}
\usepackage[center,small]{caption}
\usepackage{pstricks}
\usepackage{booktabs}
\usepackage{multicol}
\usepackage{lipsum}
\usepackage{dblfloatfix}
\usepackage{algorithm} 
\usepackage[noend]{algpseudocode} 
\usepackage[normalem]{ulem}
\usepackage{enumitem}
\usepackage{bbm}
\usepackage{dsfont}
\usepackage{subcaption}
\usepackage{comment}

\usepackage[letterpaper,%
            left=0.75in,right=0.75in,top=0.65in,bottom=0.73in,%
        footskip=.00in]{geometry}

\newcommand{\mcal}{\mathcal}
\newcommand{\msf}{\mathsf}

\newcommand{\hlambda}{\hat{\lambda}}
\newcommand{\mbb}{\mathbb}

\newtheorem{thm}{Theorem}

\newtheorem{defin}{Definition}
\newtheorem*{defin*}{Definition}
\newtheorem{prope}{Property}
\newcommand{\1}{\mathds{1}}
\algnewcommand\algorithmicforeach{\textbf{for each}}

\algdef{S}[FOR]{ForEach}[1]{\algorithmicforeach\ #1\ \algorithmicdo}

\DeclareMathOperator*{\symdiff}{\text{\Large $\Delta$}}

\input{widebar_hack.tex}

\begin{document}
\title{Polynomial-time Capacity Calculation and Scheduling for Half-Duplex 1-2-1 Networks} 
\author{
\IEEEauthorblockN{Yahya H. Ezzeldin$^\dagger$, Martina Cardone$^{\star}$, Christina Fragouli$^{\dagger}$, Giuseppe Caire$^*$}
$^{\dagger}$ UCLA, Los Angeles, CA 90095, USA,
Email: \{yahya.ezzeldin, christina.fragouli\}@ucla.edu\\
$^{\star}$ University of Minnesota, Minneapolis, MN 55404, USA,
Email: cardo089@umn.edu\\
$^*$ Technische  Universit\"{a}t  Berlin,
Berlin, Germany, 
Email: caire@tu-berlin.de
}
\IEEEoverridecommandlockouts
\maketitle
\thispagestyle{empty}
\begin{abstract}
This paper studies the 1-2-1 half-duplex network model, where two half-duplex nodes can communicate only if they point ``beams'' at each other; otherwise, no signal can be exchanged or interference can be generated. 
The main result of this paper is the design of two polynomial-time algorithms that: (i) compute the approximate capacity of the 1-2-1 half-duplex network and, (ii) find the network schedule optimal for the approximate capacity.
The paper starts by expressing the approximate capacity as a linear program with an exponential number of constraints. A core technical component consists of building a polynomial-time separation oracle for this linear program, by using algorithmic tools such as perfect matching polytopes and Gomory-Hu trees.
\end{abstract}

\section{Introduction}
Millimeter wave (mmWave) communication is expected to play an integral role in the Fifth Generation~(5G) cellular technology, and to be used in a range of services, such as ultra-high resolution video streaming, vehicle-to-vehicle communication and intelligent buildings broadband coverage~\cite{alliance20155g}. However, although a wide spectrum of promising applications has emerged fast, the focus of theoretical work has been mostly limited to single-hop systems. In this paper, we contribute to the fundamental understanding of networks of mmWave nodes building on our previous results in~\cite{EzzeldinISIT2018}.



In~\cite{EzzeldinISIT2018}, we recently introduced Gaussian 1-2-1 networks, a model that abstracts the directivity of mmWave beamforming.
Using this model, we studied the Shannon capacity for arbitrary network topologies that consist of Full-Duplex (FD) mmWave nodes, that is, nodes that can receive and transmit at the same time using two highly directive beams. In particular, in~\cite{EzzeldinISIT2018} we showed that the capacity of a Gaussian 1-2-1 network with FD mmWave nodes can be approximated to within a universal constant gap\footnote{Constant gap refers to a quantity that is independent of the channel coefficients and operating SNR, and solely depends on the number of nodes.} in polynomial-time in the network size. We term such approximation as {\em approximate capacity} in the remainder of the paper.

The focus of this paper is on Gaussian 1-2-1 networks with Half-Duplex (HD) nodes, that is, at any time instance, an HD node can either receive or transmit with a highly directive beam, but not both simultaneously. 
Our main results are as follows. We design a polynomial-time algorithm that enables to calculate the approximate capacity of a Gaussian HD 1-2-1 network with arbitrary topology. Additionally, we design an algorithm that computes the optimal schedule to achieve this {\em approximate capacity} in polynomial-time.
    \begin{figure}
        \centering
        \includegraphics[width=0.92\columnwidth]{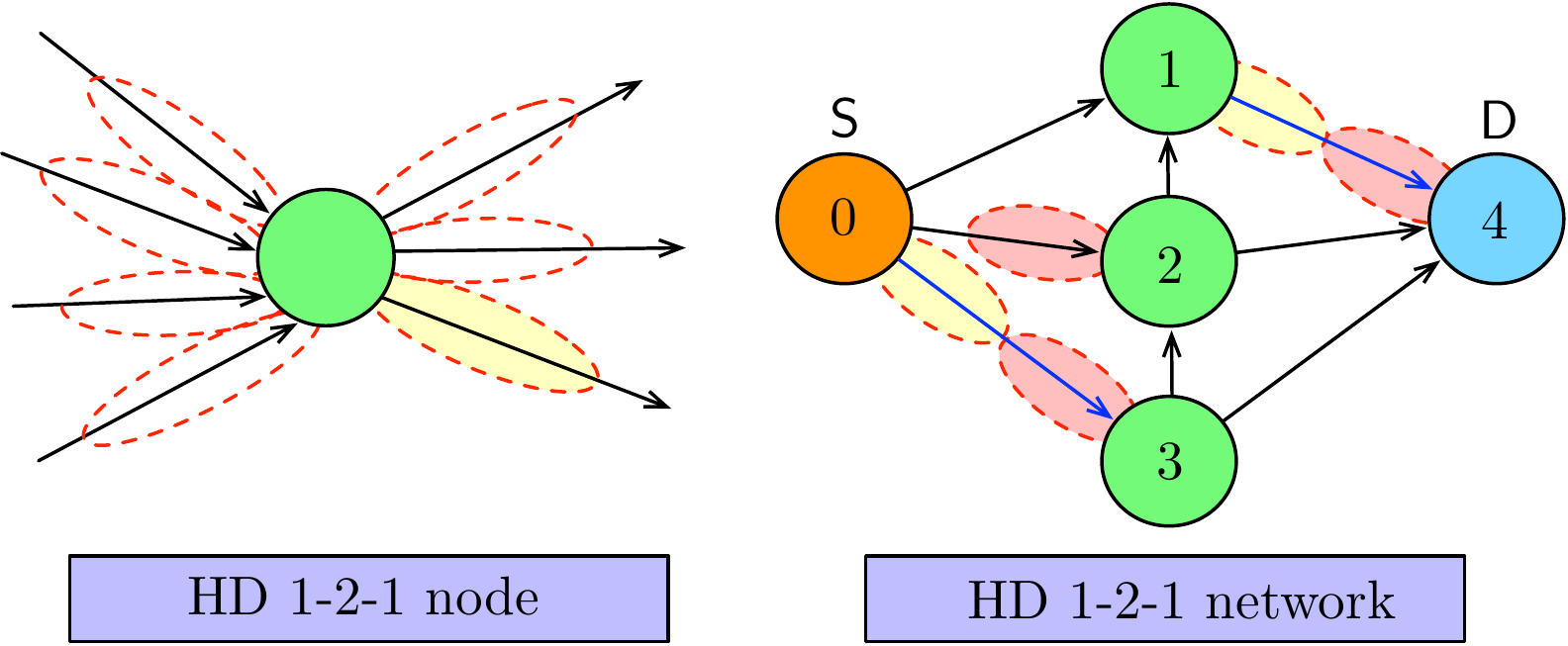}
        \caption{Model of a 1-2-1 node (left) and an example of a 1-2-1 network with $N = 3$ relays (right).}
        \vspace{-1em}
        \label{fig:example_ntwk}
    \end{figure}
This is a surprising result: the approximate capacity of HD networks is notoriously hard to study. For a network with $N$ relays, since each node can either transmit or receive, there exist $2^N$ (an exponential number of) possible states. Thus, to calculate the approximate capacity, we may need to examine the fraction of time each of these states needs to be used. For instance, for the traditional Gaussian wireless network,
although for FD networks the approximate capacity can be computed in polynomial-time in the network size~\cite{EtkinIT2014}, in HD such result is known to hold only for few special cases, such as line networks~\cite{EzzeldinISIT2017} and specific classes of layered networks~\cite{EtkinParvareshShomoronyAvestimehr}. Furthermore, although there have been several works that characterize the complexity of the optimal schedule for Gaussian HD wireless networks~\cite{brahma2016complexity,CardoneIT2014}, the problem of efficiently finding the schedule optimal for the approximate capacity for any general number of relays $N$ has only been solved for Gaussian line networks~\cite{EzzeldinISIT2017}. In this work, we show that the approximate capacity and an optimal schedule for Gaussian HD 1-2-1 networks can be {\it always} computed in polynomial-time in the network size, independently of the network topology.

Our result for HD 1-2-1 networks parallels our recent result for FD networks proved in~\cite{EzzeldinISIT2018}. However, the approach and tools we use here are different. In particular, we first show that the approximate capacity can be calculated as the solution of a linear program (LP) that has an exponential number of constraints. Then, we reduce our problem to building a polynomial-time separation oracle for this LP, by using graph-theoretic tools such as perfect matching polytopes~\cite{edmonds1965} and Gomory-Hu trees~\cite{gomory1961multi}.

\noindent{\bf{Paper Organization.}} 
Section~\ref{sec:model} describes the $N$-relay Gaussian HD 1-2-1 network and presents some known capacity results for them.
Section~\ref{sec:MainRes} presents our main results, which are proved in Section~\ref{sec:ProofOracle} and Section~\ref{sec:states_alg}.

\section{System Model and Known Results}
\label{sec:model}
With $[n_1:n_2]$ we denote the set of integers from $n_1$ to $n_2 \geq n_1$; 
$\emptyset$ is the empty set; $\1_{P}$ is the indicator function; $0^N$ is the all-zero vector of length $N$; $|S|$ is the absolute value of $S$ when $S$ is a scalar, and the cardinality when $S$ is a set.

\noindent {\bf 1-2-1 Gaussian HD network.}
We consider a network denoted by $\mcal{N}$ where $N$ relays assist the communication between a source node (node $0$) and a destination node (node $N{+}1$). We assume the following network model, which is shown in Fig.~\ref{fig:example_ntwk}.
At any time instant, this network has the two following features:
(i) each node can direct/beamform its transmissions towards at most another node,
and (ii) each node can point its receiving beam towards -- and hence receive transmissions from -- at most one other node.
Moreover, the $N$ relays operate in HD mode, that is, at any particular time, each relay can be either transmitting to or receiving from at most one node.

In particular, we can mathematically model the features explained above, by introducing two discrete state random variables $S_{i,t}$ and $S_{i,r}$, for each node $i \in [0:N+1]$. The variable $S_{i,t}$ (respectively, $S_{i,r}$) indicates the node towards which node $i$ is pointing its transmitting (respectively, receiving) beam.
With this, we have that
\begin{subequations}
\label{eq:state}
\begin{align}
&S_{i,t} \subseteq [1:N+1]  \backslash \{i \}, \ |S_{i,t}| \leq 1,
\\
&S_{i,r} \subseteq [0:N]  \backslash \{i \}, \ |S_{i,r}| \leq 1,
\\
& |S_{i,t}| \!+\! |S_{i,r}| \leq 1, \label{eq:stateHD}
\end{align}
\end{subequations}
where $S_{0,r} = S_{N+1,t} = \emptyset$ since the source always transmits and the destination always receives, and where the constraint in~\eqref{eq:stateHD} follows since the relays operate in HD, i.e., for relay $i \in [1:N]$, if $S_{i,t} \neq \emptyset$, then $S_{i,r} = \emptyset$, and vice versa.

Thus, $\forall j \in [1{:}N{+}1]$, we can write the input/output relationship for the Gaussian HD 1-2-1 network as
\begin{align}
    Y_j  = 
    \begin{cases}
        h_{j S_{j,r}} \widebar{X}_{S_{j,r}}(j) + Z_j & \text{if}\ |S_{j,r}| = 1,\\
        0 & \text{otherwise},
    \end{cases}
    \label{eq:model_2}
\end{align}
where: 
(i) $Y_j \in \mathbb{C}$ represents the channel output at node $j$;
(ii) $h_{ji} \in \mathbb{C}$ is the channel coefficient from node $i$ to node $j$; the channel coefficients are assumed to remain constant for the entire transmission duration and hence they are known by all nodes in the network;
(iii) $Z_j$ is the additive white Gaussian noise at the $j$-th node; noises across nodes in the network are assumed to be independent and identically distributed as $\mcal{CN}(0,1)$;
(iv) $\widebar{X}_i \in  \mathbb{C}^{N+1}$ has elements $\widebar{X}_i(k)$ defined as
\begin{align*}
\widebar{X}_i(k) = X_i \1_{\{k \in S_{i,t}\}},
\end{align*}
where $S_{i,t}$ is defined in~\eqref{eq:state}, and where $X_i \in \mathbb{C}$ denotes the channel input at node $i$; the channel inputs are subject to an individual power constraint, i.e., $\mathbb{E}[|X_i|^2] \leq P, \forall i \in [0:N]$; note that, when node $i$ is not transmitting, i.e., $S_{i,t} = \emptyset$, then $\widebar{X}_i = 0^{N+1}$;
(v) $S_{j,r}$ is defined in~\eqref{eq:state}.

\noindent {\bf Capacity results on 1-2-1 networks.} The Shannon capacity of the Gaussian HD 1-2-1 network described in~\eqref{eq:model_2} is not known.
However, recently we have shown in~\cite[Theorem~1]{EzzeldinISIT2018} that the capacity $\msf{C}$ of the Gaussian HD (as well as FD) 1-2-1 network can be approximated by $\msf{C}_{\rm cs,iid}$ as follows\footnote{The expression in~\eqref{eq:constGap} holds for both HD and FD with different $\mathsf{GAP}$ and different feasibility sets of the schedules $\{\lambda_s\}$.},
\begin{subequations}
\label{eq:constGap}
\begin{align}
&\msf{C}_{\rm cs,iid} \leq \mathsf{C} \leq \msf{C}_{\rm cs,iid} + \mathsf{GAP},
\\
&\msf{C}_{\rm cs,iid} \!=\! \max_{\substack{\lambda_s: \lambda_s \geq 0 \\ \sum_s \lambda_s = 1}} \min_{\Omega \subseteq [1:N] \cup \{0\}} \!\! \sum_{\substack{(i,j): i \in \Omega,\\ j \in \Omega^c}} \overbrace{\underbrace{\left( \sum_{\substack{ s:\\ j \in s_{i,t},\\ i \in s_{j,r} }} \!\! \lambda_s \right)}_{A_{ji}} \ell_{j,i}}^{\ell_{j,i}^{(s)}}, \label{eq:apprCap}
\\
& \ell_{j,i} = \log\left(1 +P\left| h_{ji}\right|^2\right),
\\
&
\mathsf{GAP} = O(N\log N)
, \label{eq:GAP}
\end{align}
\end{subequations}
where:
(i) $\Omega$ enumerates all possible cuts in the graph representing the network, such that the source belongs to $\Omega$;
(ii) $\Omega^c = [0:N+1] \backslash \Omega$;
(iii) $s$ enumerates all possible network states of the 1-2-1 network in HD (or FD), where each network state corresponds to specific values for the variables in~\eqref{eq:state} for each HD node; 
(iv) $\lambda_s$, i.e., the optimization variable, is the fraction of time for which state $s$ is active; we refer to a schedule as the collection of $\lambda_s$ for all feasible states, such that they sum up to at most 1;
(v) $s_{i,t}$ and $s_{i,r}$ denote the transmitting and receiving states for node $i$ in the network state $s$ (defined in~\eqref{eq:state} for HD operation).
In other words, for Gaussian HD 1-2-1 networks, $\msf{C}_{\rm cs,iid}$ in~\eqref{eq:constGap} is the approximate capacity of the Shannon capacity $\mathsf{C}$.

The expression in~\eqref{eq:apprCap} can be explained as follows. Given a fixed schedule $\{\lambda_s\}$, for each point-to-point link ($i {\to} j$) in the network, we sum together the activation times $\lambda_s$ of all the states $s$ that activate this link (represented by $A_{ji}$ in~\eqref{eq:apprCap}). Then, we weight/multiply the link capacity $\ell_{j,i}$ by this effective activation time $A_{ji}$. For this new network with weighted link capacities $\ell_{j,i}^{(s)}$ as shown in~\eqref{eq:apprCap}, we calculate the graph-theoretic min-cut. Finally, we maximize this min-cut over all possible feasible schedules of the 1-2-1 network.


\section{Main Results}
\label{sec:MainRes}
In this section, we show how the expression of the approximate capacity $\msf{C}_{\rm cs,iid}$ in~\eqref{eq:apprCap} for Gaussian HD 1-2-1 networks can be efficiently evaluated and how an optimal schedule $\{\lambda_s\}$ for~\eqref{eq:apprCap} can be found in polynomial-time in the number of network relays $N$. In particular, our main result is summarized by the following theorem.

\begin{figure*}
    \centering
    \begin{subfigure}[b]{0.3\textwidth}
        \includegraphics[width=\textwidth]{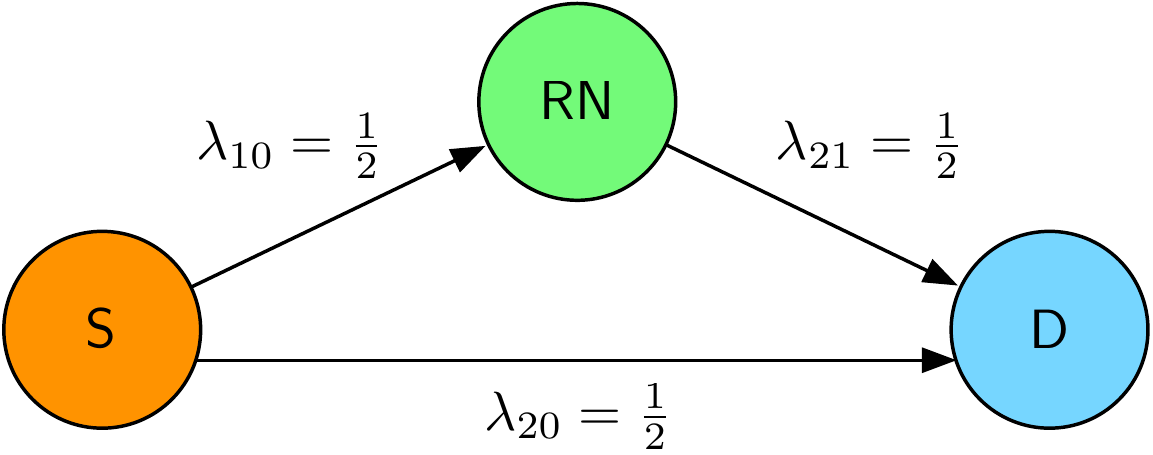}
        \caption{1-2-1 network with link activations.}
        \label{fig:NetExConst}
    \end{subfigure}
~
    \begin{subfigure}[b]{0.3\textwidth}
        \includegraphics[width=\textwidth]{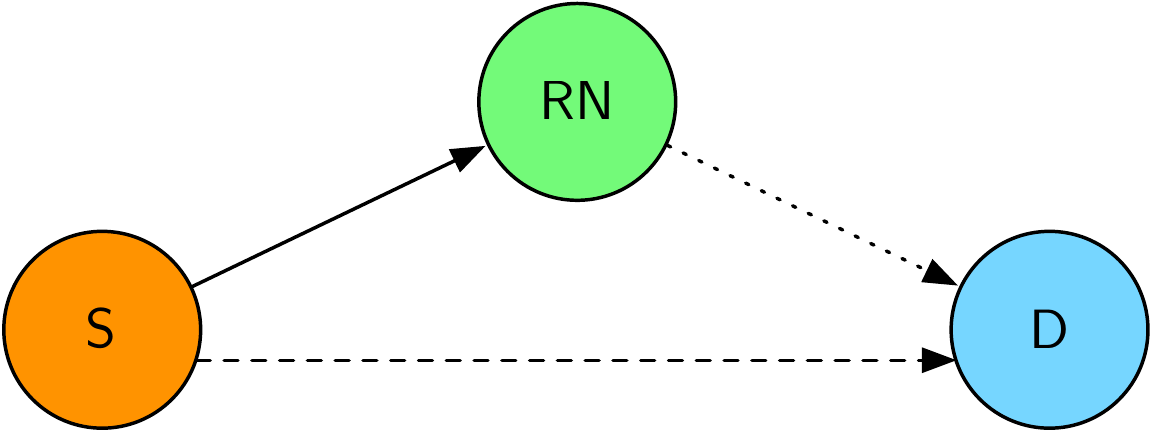}
        \caption{Network states in HD.}
        \label{fig:StatesHD}
    \end{subfigure}
~
    \begin{subfigure}[b]{0.3\textwidth}
        \includegraphics[width=\textwidth]{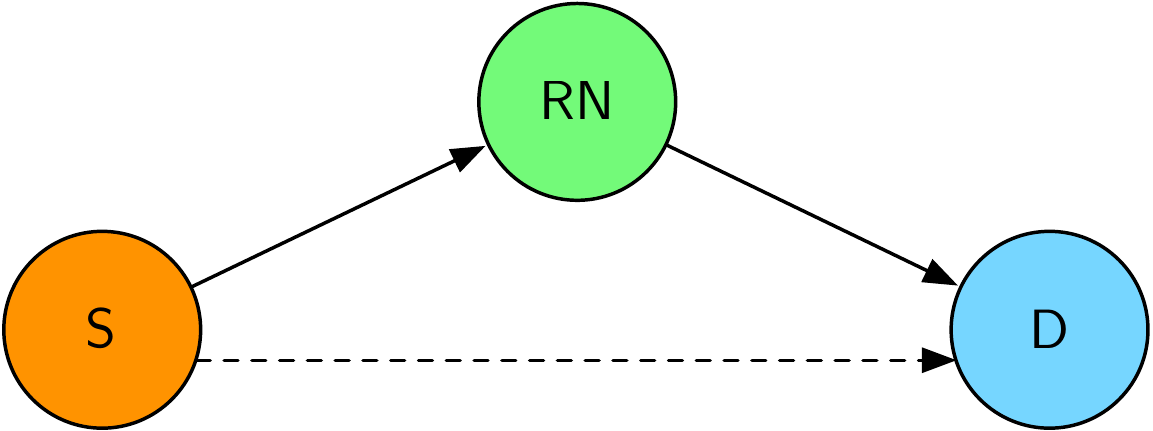}
        \caption{Network states in FD.}
        \label{fig:StatesFD}
    \end{subfigure}
    \caption{Gaussian 1-2-1 network examples and network states in HD and FD.}\label{fig:NetExTot}
\end{figure*}

\begin{thm}
\label{cor:HDCapPoly}
For the $N$-relay  Gaussian HD 1-2-1 network:
\begin{enumerate}[label=(\alph*)]
\item The approximate capacity $\msf{C}_{\rm cs,iid}$ can be found in polynomial time in $N$;
\item An optimal schedule for the approximate capacity $\msf{C}_{\rm cs,iid}$ can be found in polynomial-time in $N$.
\end{enumerate}
\end{thm}
To the best of our knowledge, Gaussian HD 1-2-1 networks represent the first class of HD relay networks for which the approximate capacity and schedule can be computed efficiently independently of the network topology. 
In what follows, we focus on the proof of Theorem~\ref{cor:HDCapPoly}.


\noindent {\bf Theorem~\ref{cor:HDCapPoly} Part (a).} The proof of the first part of Theorem~\ref{cor:HDCapPoly} is a direct consequence of two results that we present and discuss in what follows.
Our first result shows that calculating $\msf{C}_{\rm cs,iid}$ for Gaussian HD 1-2-1 networks is equivalent to solving an LP, where state activation times are replaced by link activation times. In particular, we have the following~theorem for which the proof is delegated to Appendix A.

\begin{thm}
\label{thm:ThLinkAct}
For any $N$-relay Gaussian HD 1-2-1 network, we have that
\begin{align*}
\rm{P1}&{\rm\ :} \ \msf{C}_{\rm cs,iid} = \max \sum_{j =1}^{N+1} F_{j,0}& \nonumber\\
& (1a) \quad 0 \leq F_{j,i} \leq  \lambda_{ji} \ell_{j,i}, \quad \forall (i,j) \in [0{:}N]\times[1{:}N{+}1], \nonumber\\
& (1b) \sum_{j\in[1:N{+}1]\backslash\{i\}} F_{j,i} = \sum_{k\in[0:N]\backslash\{i\}} F_{i,k}, \quad i \in [1:N],\nonumber\\
& (1c)\  \hat{\lambda}_{ij} = \lambda_{ij} + \lambda_{ji}, \quad i \in [0{:}N],\ j \in [i{+}1{:}N{+}1], \nonumber\\
& (1d)\  \lambda_{ij} \geq 0 \quad (i,j) \in [0{:}N]\times[1{:}N{+}1], \\
& (1e)\  \sum_{\substack{(i,j) : i = v\ {\rm or}\ j = v,\\ i < j}} \hat{\lambda}_{ij} \leq 1, \quad \forall v \in [0:N+1], \nonumber\\
& (1f)  \! \! \!\sum_{\substack{i \in S, j \in S,\\ i < j}} \!\!\!\!\hat{\lambda}_{ij} \! \leq \! \frac{|S| - 1}{2}, \ \forall S \! \subseteq \! [0:N\!+\!1],\ |S|\ {\rm odd},\nonumber\\
& (1g)\  \hat\lambda_{ij} \geq 0 \quad i \in [0{:}N],\ j \in [i{+}1{:}N{+}1], \nonumber
\end{align*}
where $F_{j,i}$ represents the data flow through the link of capacity $\ell_{j,i}$ and $\lambda_{ji}$ represents the fraction of time in which the link is active.
\end{thm}

Note that $\rm{P1}$ is very similar to the LP representation of the max-flow problem, where the edge capacities are given by $\lambda_{ji} \ell_{j,i}$. The key difference is that $\lambda_{ji}$ is now a variable and is subject to the feasibility constraints in $(1c) - (1f)$ that stem from the nature of the scheduling in HD 1-2-1 networks. Note that a similar LP as in Theorem~\ref{thm:ThLinkAct} was obtained in~\cite{EzzeldinISIT2018} for the FD case. However, there is a fundamental difference in HD, which is captured by the constraints in $(1f)$ that are not needed in FD. 
To illustrate the need of the constraint in $(1f)$ in $\rm{P1}$, consider the network in Fig.~\ref{fig:NetExConst}.
Assume that each of the three links in the network is active for a fraction of time equal to $1/2$ (as shown in Fig.~\ref{fig:NetExConst}).
Clearly, these link activation times satisfy the constraint in $(1c){-}(1e)$ with
$\hat{\lambda}_{01} {=} \hat{\lambda}_{02} {=} \hat{\lambda}_{12} = \frac{1}{2}$.
However, we note that these link activation times do not satisfy the constraint in $(1f)$ since, by considering $S = {\{}0{,}1{,}2{\}}$, we have
\begin{align*}
\sum_{\substack{i \in S, j \in S,\ i < j}} \!\!\!\!\hat{\lambda}_{ij} = \hat{\lambda}_{01} + \hat{\lambda}_{02} + \hat{\lambda}_{12} = \frac{3}{2} > \frac{|S| - 1}{2} = 1.
\end{align*}
Thus, if the constraint in $(1f)$ was not there, then one would conclude that the link activation times illustrated in Fig.~\ref{fig:NetExConst} are feasible. 
However, we now show that this is not the case, which highlights the need of the constraint in $(1f)$.
For the Gaussian 1-2-1 network in Fig.~\ref{fig:NetExConst}, when the relay operates in HD, there are three possible useful states of the network, in each of which exactly one link is active. These three states are depicted with different line styles (i.e., solid, dashed, dotted) in Fig.~\ref{fig:StatesHD}.
Note that the links $\msf{S}{\to}\msf{RN}$ and $\msf{RN}{\to}\msf{D}$
cannot be active simultaneously because of the HD constraint at the relay.
Additionally, the links $\msf{S}{\to}\msf{RN}$ and $\msf{S}{\to}\msf{D}$ cannot be activated simultaneously since the source has only a single transmitting beam. 
A similar argument also holds for the links $\msf{RN}{\to}\msf{D}$ and $\msf{S}{\to}\msf{D}$.
Thus, for this network we have a one-to-one mapping between $\lambda_s$ in~\eqref{eq:apprCap} and $\lambda_{ji}$ in $\rm{P1}$, with $\lambda_s=\lambda_{ji}$ if state $s$ activates the link of capacity $\ell_{j,i}$. Hence, if we use the values from Fig.~\ref{fig:NetExConst}, we would obtain $\sum_s \lambda_s = 3/2 > 1$
which clearly does not satisfy the constraint in~\eqref{eq:apprCap}.
We therefore conclude that the link activation times illustrated in Fig.~\ref{fig:NetExConst}, which are feasible if the relay operates in FD (see Fig~\ref{fig:StatesFD}), are not feasible when the relay operates in HD.
This simple example shows why the constraint in $(1f)$ in $\rm{P1}$ is needed for HD Gaussian 1-2-1 networks.

We note that the LP $\rm{P1}$ has a number of variables that is polynomial in $N$ (two per each edge in the network) compared to the number of variables in the maximization in~\eqref{eq:apprCap}, which instead is exponential in $N$ (one per each state in the network). However, we also note that $\rm{P1}$ now has an exponential number of constraints of the type $(1f)$.
Thus, it follows that algorithms such as the simplex method and the interior point method can not solve $\rm{P1}$ in polynomial-time in $N$.
However, as we show next, the ellipsoid method~\cite{grotschel1981ellipsoid} can indeed solve $\rm{P1}$ in polynomial-time in $N$.
The key step of the ellipsoid method, that incorporates the constraints of an LP, relies on the existence of an oracle which, given the problem and a point in space, can decide in polynomial-time whether the point is feasible or not and, if not, it returns one constraint of the linear program that is violated by that point. This is referred to as a {\em{polynomial-time separation oracle}}. 
Our next result focuses on showing that a polynomial-time separation oracle for $\rm{P1}$ exists such that, given the graph representing the $N$-relay Gaussian HD 1-2-1 network and a point $y$ in the space of $\rm{P1}$, it can verify in polynomial-time in $N$ if $y$ is feasible in $\rm{P1}$ and, if not feasible, it returns one of the constraints that is violated. In other words, if one of the constraints is violated, then the oracle returns a hyperplane that separates the point $y$ from the feasible polytope in $\rm{P1}$.
This result is formalized in the theorem below for which the proof is given in Section~\ref{sec:ProofOracle}.

\begin{thm}
\label{thm:Oracle}
A polynomial-time separation oracle exists that, provided with a weighted graph with $N+2$ nodes and a point $y$ in the space of $\rm{P1}$, can verify in polynomial-time in $N$ if $y$ is feasible in $\rm{P1}$, and if not feasible it returns one of the constraints in $\rm{P1}$ that is violated.
\end{thm}
Theorem~\ref{thm:ThLinkAct}, Theorem~\ref{thm:Oracle} and the existence of the ellipsoid method~\cite{grotschel1981ellipsoid} directly imply the result in part (a) of Theorem~\ref{cor:HDCapPoly}. 

\noindent {\bf Theorem~\ref{cor:HDCapPoly} Part (b).} The proof of the second part of Theorem~\ref{cor:HDCapPoly} makes use of Theorem~\ref{thm:Oracle} and an algorithmic version of Caratheodory's theorem to find a feasible schedule $\lambda_s$ for the approximate capacity in~\eqref{eq:apprCap}, such that each link is activated
for the amount given by the solution of $\rm{P1}$. Theorem~\ref{cor:HDCapPoly} Part (b) is proved in Section~\ref{sec:states_alg}. 

\section{Proof of Theorem~\ref{thm:Oracle}}
\label{sec:ProofOracle}
In this section, we prove Theorem~\ref{thm:Oracle}, namely we show the existence of a polynomial time separation oracle that, provided with a weighted graph with $N+2$ nodes -- representing our Gaussian HD 1-2-1 network -- and a point $y \!=\! (F_{i,j},\lambda_{ij}, \hat{\lambda}_{ij})$ in the space of $\rm{P1}$, can verify in polynomial time in $N$ if $y$ is feasible in $\rm{P1}$ and if not, it returns a hyperplane that separates $y$ from the feasible region (i.e., an inequality satisfied for the feasible region but not for $y$).

Our oracle can be divided into two parts: (i) a simple oracle that checks the constraints in $(1a)$-$(1e)$, and (ii) a more involved oracle for checking the constraint in $(1f)$.
Note that since the number of variables and constraints in $(1a)$-$(1e)$ is polynomial in $N$, then we can directly check these constraints for $y$ in polynomial time in $N$. 
If one constraint is violated, then we return that constraint as the hyperplane that separates $y$ from the feasible set.
In what follows, we prove that the constraint in $(1f)$ can also be checked in polynomial time in $N$.
Towards this end, in Section~\ref{subsec:MpolPMpol} we first overview some results from~\cite{edmonds1965}, and define the M-polytope and the Perfect Matching polytope (PM-polytope) of an undirected graph, and show a useful relationship between them. Note that (similar to the definition of the M-polytope), the PM-polytope of an undirected graph $G$ is the polytope that has all perfect-matchings\footnote{A perfect matching is a matching such that all vertices in the graph are connected to one edge in the matching set.} as its extreme points.
Then, in Section~\ref{subsec:GHTree} we show how a set $S$ that violates the constraint in $(1f)$ can be found by first constructing a Gomory-Hu tree~\cite{gomory1961multi} of the weighted graph representing our network, and then checking cuts with a particular structure in it.
Finally, in Section~\ref{subsec:polyOr}, we show how these results can be leveraged to build our polynomial time separation oracle.

\subsection{M-polytope and PM-polytope}
\label{subsec:MpolPMpol}
We here overview some results from~\cite{edmonds1965}, and show a useful relationship between the M-polytope and the PM-polytope of an undirected graph, which we next define.
In particular, we will use the following graph theory notation.

For an undirected graph $G = (V_G,E_G,x_G)$, with set of vertices $V_G$, set of edges $E_G$ and edge weight function $x_G : E_G \to \mathbb{R}$, we use the convention $e = (i,j)$ with $i < j$.
Furthermore, in the remainder of this section, we use the definitions in Table~\ref{table:Notation}.

\begin{table*}
\caption{Quantities of interest used throughout Section~\ref{sec:ProofOracle}.}
\label{table:Notation}
\begin{center}
\begin{tabular}{ |c||c| }
 \hline
 {\bf{Quantity}} & {\bf{Definition}}   \\
 \hline
\hline
$x_G(e)$ & Weight corresponding to edge $e = (i,j)$ between nodes $i \in V_G$ and $j \in V_G$ \\ \hline
$\delta_G(S)$ & Set of edges in $G$ that have only one endpoint in the set of vertices $S \subseteq {V_G}$ \\ \hline
$E_G(A,B)$ & Set of edges with one endpoint in the set of vertices $A$ and the other endpoint in $B$ with  $A,B \subseteq {V_G}$ \\ \hline
$E_G(A)$ & Set of edges with both endpoints in the set of vertices $A$ \\ \hline
$x_G(F)$ & Sum of the weights of the edges that belong to $F$, i.e., $x_G(F) = \displaystyle\sum_{e \in F} x_G(e)$ \\
 \hline
\end{tabular}
\end{center}
\vspace{-6mm}
\end{table*}

\begin{figure}
\centering
\includegraphics[width=\columnwidth]{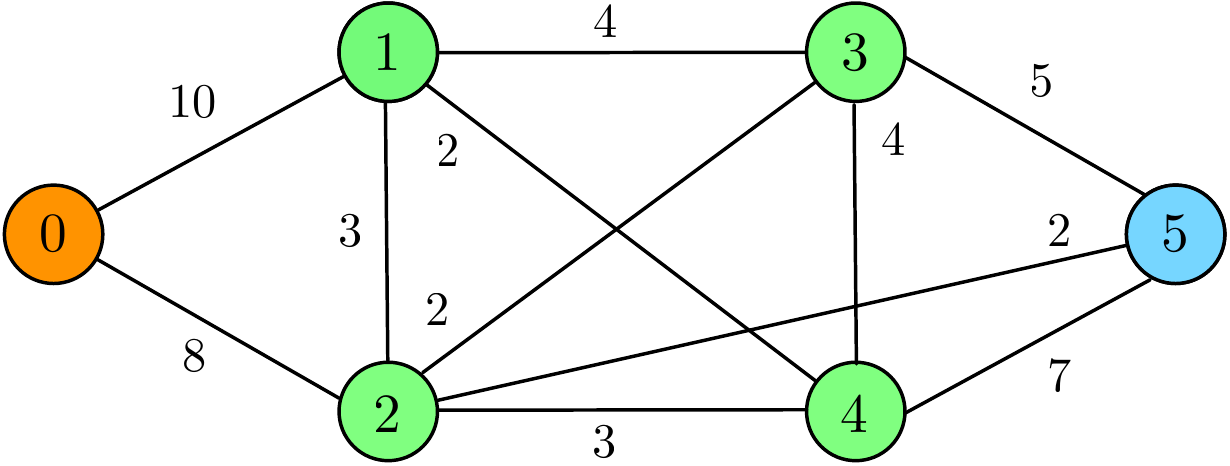}
\caption{Example of a weighted graph $G$.}
\label{fig:GraphEx}
\end{figure}

As an example of the used notation,
with reference to the weighted graph $G$ in Fig.~\ref{fig:GraphEx}, we have
\begin{align*}
&\delta_G(\{1,3,4,5\}) = \{(0,1),(1,2),(2,3),(2,4),(2,5)\},
\\
&E_G(\{1,3,4,5\},\{2\})  = \{(1,2),(2,3),(2,4),(2,5)\},
\\
&E_G(\{1,3,4,5\})  = \{(1,3),(1,4),(3,4),(3,5),(4,5)\},
\\
&x_G(\{(0,2), (1,3),(2,5)\})  = 8 + 4 + 2 = 14.
\end{align*}
We let $G = (V_G,E_G,\hlambda)$ define the weighted undirected graph that describes our Gaussian HD 1-2-1 network, where $\hlambda(e) = \hlambda_e$ is defined as in the constraint in $(1c)$ in $\rm{P1}$. 
Then, we note that the constraints in $(1d)-(1f)$ in $\rm{P1}$ are those introduced by Edmonds~\cite{edmonds1965} to define the M-polytope for the graph $G$.
By using the notation introduced above, we can rewrite the constraints in $(1d)-(1f)$ in $\rm{P1}$, as follows
\begin{align}
\label{eq:MP}
\begin{array}{ll}
   	{\rm M-polytope} :  & \\
\qquad \hlambda(e) \geq 0,& \forall e \in E_G,\\
   	\qquad \hlambda(\delta_G(v)) \leq 1,& \forall v \in V_G,\\
   	\qquad \hlambda(E_G(S)) 
   	\leq \dfrac{|S|-1}{2},& \forall S \subseteq V_G, |S| \ \text{odd}.
   	\end{array}
\end{align}
The PM-polytope is represented similarly to the M-polytope, where now the second constraint in~\eqref{eq:MP} is forced to be satisfied with equality, namely 
\begin{align}\label{eq:PMP_}
\begin{array}{ll}
   	{\rm PM-polytope} : &  \\
\qquad  \hlambda(e) \geq 0,& \forall e \in E_G,\\
 \qquad  	\hlambda(\delta_G(v)) = 1,& \forall v \in V_G,\\
\qquad   	\hlambda(E_G(S)) \leq \dfrac{|S|-1}{2},& \forall S \subseteq V_G, |S| \ \text{odd}.
   	\end{array}
\end{align}
By using the second constraint in~\eqref{eq:PMP_}, we can manipulate the third constraint, and rewrite the PM-polytope -- see Appendix~\ref{sec:PM_equiv} for the detailed computation -- as
\begin{align}\label{eq:PMP}
\begin{array}{ll}
   	{\rm PM-polytope} : & 
\\ \qquad  \hlambda(e) \geq 0,& \forall e \in E_G,\\
  \qquad  	\hlambda(\delta_G(v)) = 1,& \forall v \in V_G,\\
   \qquad 	\hlambda(\delta_G(S)) \geq  1,& \forall S \subseteq V_G, |S| \ \text{odd}.
   	\end{array}
\end{align}
We now show that, for any graph $G$ with $N$ vertices, there is an injection from the M-polytope of $G$ to the PM-polytope of a constructed graph $\widetilde{G}$ with double the number of vertices.
Towards this end, we first create a copy $G' = (V_{G'}, E_{G'},\hlambda')$ of the original graph $G$ and we let $\widetilde{G} = (V_{\widetilde{G}}, E_{\widetilde{G}},\widetilde{\lambda})$ be a graph with: 
(i) $V_{\widetilde{G}} = V_G \cup V_{G'}$,
(ii) $E_{\widetilde{G}} = E_{G} \cup E_{G'} \cup \{(v,v') | v \in V_G,\ v' \ \text{is the copy of }v \text{ in } G'  \}$, and
(iii) $\widetilde{\lambda}(e)$ defined as
\begin{align}
\label{eq:lambdaTilde}
   \widetilde{\lambda}(e) =
   \begin{cases}
      \hlambda(e) & \text{if } e \in E_G,\\
      \hlambda'(e) & \text{if } e \in E_{G'},\\
      1 - \hlambda(\delta_G(v)) & \text{if } e = (v,v').
   \end{cases}
\end{align}
We now show that, if $\hlambda$ is in the M-polytope of $G$, then $\widetilde\lambda$ is in the PM-polytope of $\widetilde G$.
We start by noting that the given construction for $\widetilde\lambda$ in~\eqref{eq:lambdaTilde} satisfies the first two constraints in~\eqref{eq:PMP}. 
The key challenge is to show that the third constraint is also satisfied, i.e.,
\begin{align}\label{eq:perfect_graph}
	\widetilde\lambda(\delta_{\widetilde{G}}(S)) \geq  1,\quad  \forall S \subseteq V_{\widetilde{G}},\ |S| \ \text{odd}.
\end{align}
In Appendix~\ref{app:RelPMandM}, we show that the constraint in~\eqref{eq:perfect_graph} is indeed also satisfied.
This result implies that we can check whether $\hlambda(e)$ is in the M-polytope of $G$ by checking whether $\widetilde \lambda(e)$, with the construction in~\eqref{eq:lambdaTilde}, is in the PM-polytope of $\widetilde G$.

\subsection{PM-polytope and Gomory-Hu tree}
\label{subsec:GHTree}

In the previous subsection, we have defined the PM-polytope for the weighted undirected graph ${\widetilde G}$ as in~\eqref{eq:PMP}.
In particular, in~\eqref{eq:PMP} we can check in polynomial time if the first two sets of constraints are satisfied.
Therefore, our main concern lies in the third group of constraints, since there is an exponential number of them.
We here show that a set $S$ that violates the third constraint in~\eqref{eq:PMP} can be found by first constructing a Gomori-Hu tree~\cite{gomory1961multi} of ${\widetilde G}$ (defined below), and then checking cuts with a particular structure in it.

We start by noting that the third group of constraints in~\eqref{eq:PMP} can be written in a compact way as 
\begin{align}\label{eq:odd_cut}
\min_{\substack{S \subseteq V_{\widetilde G},\\ |S| \ \text{odd}}} \hlambda(\delta_{\widetilde G}(S)) \geq  1.
\end{align}
In words, what this says is that the {\it minimum odd} cut in ${\widetilde G}$ has a value greater than or equal to 1. An {\it odd} cut is a vertex partition of $V_{\widetilde G}$ into $S$ and $S^c$ such that either $S$ and/or $S^c$ has an odd cardinality ($S^c$ is the complement of $S$).

In~\cite{padberg1982odd}, Padberg and Rao provided an efficient algorithm to find the minimum odd cut and its value for any graph $G$. 
An appealing feature of the algorithm designed in~\cite{padberg1982odd} is that it runs in polynomial time in $|V_{G}|$.
In particular, the method introduced consists of using {\it Gomory-Hu} trees.

\begin{defin}[Gomory-Hu Tree]\rm 
Let $G = (V_G,E_G,x_G)$ be a capacitated (weighted) undirected graph with capacity function $x_G : E_G \to \mbb{R}_+$.
A {\it Gomory-Hu} tree (for $G$ and $x_G$) is a capacitated tree $T = (V_G,F,\alpha_T)$ with capacity $\alpha_T$, such that for each edge $e = (s,t) \in F$, the two components of $T\backslash e$ give a a minimum capacity $s-t$ cut in $G$. The capacity of the cut in $G$ is equal to $\alpha_T(s,t)$.
\label{def:GHTree}
\end{defin}

Note that, for any capacitated undirected graph with $N$ vertices, a Gomory-Hu tree always exists and can be constructed by the algorithm in~\cite{gomory1961multi} using $|V_G|-1$ runs of the max-flow problem. Given the triangle inequality of min-cuts in a graph, Definition~\ref{def:GHTree} implies the following property of Gomory-Hu trees.

\begin{prope}
\label{prop:GHTree}
\rm 
Let $G = (V_G,E_G,x_G)$ be a capacitated undirected graph and $T = (V_G,F,\alpha_T)$ be a Gomory-Hu tree of $G$. Consider any two vertices $u,v \in V_G$, let $\mcal{P}_{uv}$ be the path connecting $u$ and $v$ in $T$ and let $(s,t)$ be the edge with the minimum capacity $\alpha_T(s,t)$ along the path $\mcal{P}_{uv}$. Then, we have the two following properties:
\begin{enumerate}
	\item The two components $T\backslash(s,t)$ give a minimum capacity $u-v$ cut in $G$;
	\item The value of the minimum capacity $u-v$ cut is given by $\alpha_T(s,t)$.
\end{enumerate}
\end{prope}

\begin{figure}
\centering
\includegraphics[width=\columnwidth]{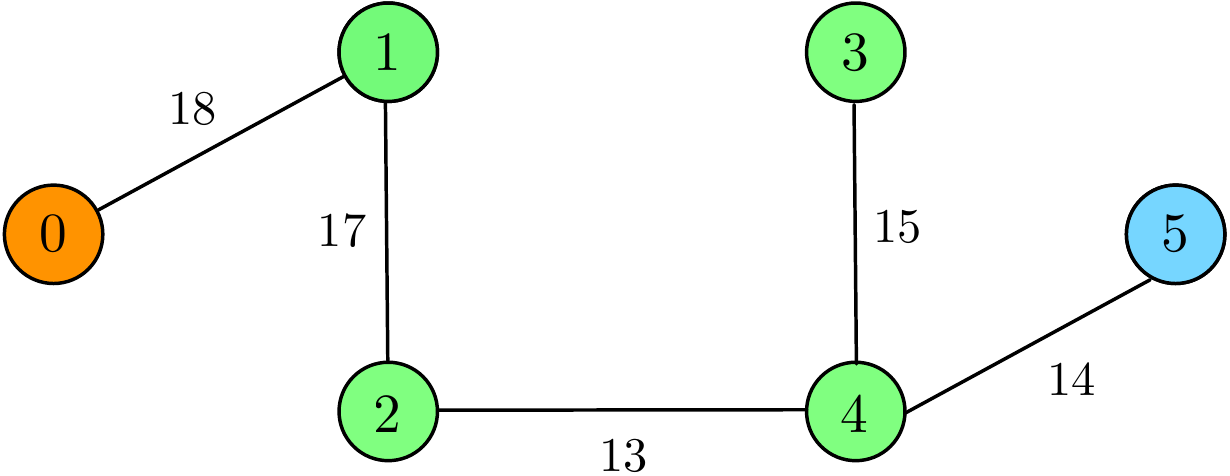}
\caption{A Gomory-Hu tree $T$ of the graph $G$ in Fig.~\ref{fig:GraphEx}.}
\label{fig:gomory_hu_example}
\end{figure}

As an example of a Gomory-Hu tree, consider Fig.~\ref{fig:gomory_hu_example}, which represents a Gomory-Hu tree $T$ of the graph $G$ in Fig.~\ref{fig:GraphEx}.
From Property~\ref{prop:GHTree}, it follows that if for $G$ in Fig.~\ref{fig:GraphEx}, we would like to know the minimum cut between vertices $1$ and $5$, then all we need to do is to look at the unique path connecting $1$ and $5$ in $T$ in Fig.~\ref{fig:gomory_hu_example}. 
The edge with the minimum capacity is $(2,4)$ and hence the min-cut between $1$ and $5$ equals $13$. 
A cut that has this value in $G$ is the partition $U = \{0,1,2\}$ and $U^c = \{3,4,5\}$.

We now state a simplified version of the result proved in~\cite{padberg1982odd} that shows how a Gomory-Hu tree of a graph can be leveraged to verify if the minimum odd cut of a graph satisfies~\eqref{eq:odd_cut}. 
This theorem is proved in Appendix~\ref{sec:odd_cuts_thm}.

\begin{thm}\label{thm:odd_cuts}
\rm 
Let ${\widetilde G} = (V_{\widetilde G},E_{\widetilde G},\widetilde\lambda)$ be an undirected capacitated graph with $|V_{\widetilde G}|$ even, and let $T = (V_{\widetilde G},F,\alpha_T)$ be a Gomory-Hu tree for ${\widetilde G}$. Then, one of the cuts determined by $T\backslash e$, $\forall e \in F$, is a minimum capacity odd cut in ${\widetilde G}$.
\end{thm}

Thus, we can use the result in Theorem~\ref{thm:odd_cuts} to verify whether the minimum odd cut has a weight greater than or equal to one, by following the procedure illustrated in Algorithm~\ref{Algo:MinCut}.
We note that in Algorithm~\ref{Algo:MinCut}, even though we need only one among $|W_f|$ and $|W_f^c|$ to be odd, in a graph with even number of vertices, if one is odd, then also the other is odd.
This explains why we check if both $|W_f|$ and $|W_f^c|$ are odd.

In summary, we have here shown that we can find a set $S$ (if any) that violates the third constraint of the PM-polytope of ${\widetilde G} = (V_{\widetilde G},E_{\widetilde G},\widetilde\lambda)$ defined in~\eqref{eq:PMP} by first constructing a Gomory-Hu tree $T = (V_{\widetilde G},F,\alpha_T)$ and then checking odd cuts in it.
Since the number of cuts in $T$ is $|V_{\widetilde G}|-1 = 2N+3$, then by using this procedure we only need to perform $O(N)$ checks (and not an exponential number of them).

\begin{algorithm}
  \caption{Check if minimum odd cut in ${\widetilde G}$ satisfies~\eqref{eq:odd_cut}}
  \label{Algo:MinCut}
  \begin{algorithmic}[1]
    \Function{CheckMinimumOddCut}{${\widetilde G}$}
      \State Build a Gomory-Hu tree $T = (V_{\widetilde G},F,\alpha_T)$ of ${\widetilde G}$
	\ForEach{$f \in F$}
\State Let $W_f$ and $W_f^c$ be the two components of $T\backslash f$
      \If{$|W_f|$ and $|W_f^c|$ are odd}
        \If{$\alpha_T(f)<1$}
\State \Return $W_f$ as the set that violates~\eqref{eq:odd_cut} 
	\EndIf
      \EndIf
\EndFor
      \Return $W_f = \emptyset$
    \EndFunction
  \end{algorithmic}
\end{algorithm}

\subsection{Polynomial time separation oracle}
\label{subsec:polyOr}
We here show how the results that we discussed and stated in the previous two subsections can be combined and leveraged to build our polynomial time separation oracle.
In particular, Algorithm~\ref{algo:PolyTimeOr} provides the pseudocode of our separation oracle.
It is worth noting that each step in Algorithm~\ref{algo:PolyTimeOr} can be performed in polynomial time in $N$ and hence our oracle runs in polynomial time in $N$. 
In particular, the complexity is dominated by the step where a Gomory-Hu tree is constructed for ${\widetilde G}$. 
This construction algorithm, in fact, involves performing the max-flow problem $2N + 3$ times and hence the complexity of this step -- and consequentially of our oracle -- is $O(N^4)$ in the worst-case.
This concludes the proof of Theorem~\ref{thm:Oracle}.

\begin{algorithm}
  \caption{Polynomial time separation oracle for $\rm{P1}$}
\label{algo:PolyTimeOr}
  \begin{algorithmic}[1]
  	\renewcommand{\algorithmicrequire}{\textbf{Input:}}
  	\renewcommand{\algorithmicensure}{\textbf{Output:}}
  	\Require Network NET, point $y$ to test for feasibility
  	\Ensure Feasible flag, violated constraint in $\rm{P1}$
  	\medskip
  	\State Check if all constraints in $(1a)$-$(1e)$ in $\rm{P1}$ are satisfied
  	\If {one constraint in $(1a)$-$(1e)$ is violated}
  		\State \Return Feasible = False, constraint violated
  	\EndIf
  	\State Construct undirected simple graph $G = (V_G ,E_G ,\hlambda)$ with the same set of nodes in NET, edges representing links
in NET (but only one direction) and $\hlambda$ given by $y$
	\smallskip
	\State Construct undirected graph $\widetilde G$ from $G$ that has double number of vertices as described in Section~\ref{subsec:MpolPMpol}
	\smallskip
	\State $W_f$ = \Call{CheckMinimumOddCut}{$\widetilde G$}
  	\If{$W_f = \emptyset$}
  		\State \Return Feasible = True
  	\Else
  		\State Feasible = False
  		\State Let $W_{f,a} = W_f \cap V_G$ and $W_{f,b}' = W_f \cap V_{G'}$
  		\State Let $W_{f,a}'$ be a copy of $W_{f,a}$ in $V_{G'}$
  		\State Let $W_{f,b}$ be a copy of $W_{f,b}'$ in $V_G$  
  		\If{$|W_{f,a}\backslash W_{f,b}|$ odd}
  		\State $Z = W_{f,a}\backslash W_{f,b}$
  		\Else 
  			\State $Z = W_{f,b}'\backslash W_{f,a}'$
  		\EndIf
  
  		\State \Return Feasible = False, constraint $Z$ in $(1f)$ violated
  	\EndIf
  \end{algorithmic}
\end{algorithm}

\section{Finding a schedule in polynomial-time}\label{sec:states_alg}
To prove part (b) of Theorem~\ref{cor:HDCapPoly}, we first note that, for a Gaussian HD 1-2-1 network, a state $s$ in~\eqref{eq:apprCap} does not activate two adjacent links.
Thus, a state is a matching of directed edges in a directed graph representing the network topology.
Now, assume that we are given a feasible point in the LP $\rm{P1}$ (obtained by solving $\rm{P1}$ as described in Section~\ref{sec:ProofOracle}). The main objective of this section is to efficiently construct a set of matchings (representing states in the network) and find their corresponding activation times (representing $\lambda_s$ in \eqref{eq:apprCap}), such that the fraction of time a link $i {\to} j$ is active is equal to $\lambda_{ji}$ in $\rm{P1}$.
For any pair of nodes $i < j$, we refer to $\hat\lambda_{ji}$ in $\rm{P1}$ as the {\em connection} activation time, i.e., $\hat\lambda_{ji}$ represents the duration of time nodes $i$ and $j$ are connected, without considering the direction of communication between them. Thus, from a connection activation time perspective, the network is represented by an undirected graph where an edge $(i,j)$ is active for a fraction $\hat \lambda_{ji}$ of time.
We first discuss how we can decompose the connection activation times into undirected matchings, and then show how these can be leveraged to construct our set of directed matchings (states). 
The goal is to show that both these tasks can be performed in polynomial-time in the number of relays $N$.

\subsection{Decomposition into undirected matchings}
We define the undirected graph $G = (V_G,E_G,\hat\lambda)$, where: (i) the graph vertices in $V_G$ represent the nodes in our Gaussian HD 1-2-1 network, (ii) $E_G = \{(i,j) | i > j,\ \hat\lambda_{ij} > 0\}$ is the set of edges, and (iii) the edge weights are equal to the values of $\hat\lambda_{ij}$ from the feasible point in $\rm{P1}$. Note that, without loss of generality, in the definition of $E_G$ we do not include any edge $e$ for which $\hat \lambda_e = 0$.

Let $\hat\lambda \in \mbb{R}^{|E_G|}$ be the vector comprised of $\hat \lambda_e,\ \forall e \in E_G$.
As highlighted in Section~\ref{sec:ProofOracle},
the constraints on $\{\hat\lambda_{e}\}$ in $(1e)-(1g)$ describe the M-polytope of the undirected graph $G$~\cite{edmonds1965}. 
Our goal here is to efficiently find a set of $K$ matchings ${M_k} \in \mbb{R}^{|E_G|}$ (vertices of the M-polytope) such that
\begin{align}\label{eq:carathedory_decomp}
	\hat \lambda = \sum_{k=1}^K \varphi_k M_k,\quad \sum_{k=1}^K \varphi = 1,\quad \varphi_k \geq 0,\ \forall k \in [1{:}K].
\end{align}
By Caratheodory's theorem~\cite{barany1982generalization}, we know that for some $K {\leq} |E_G| {+} 1$, such a decomposition of $\hat \lambda$ exists. 
However, the key challenge is to discover this decomposition in polynomial-time in $N$.
Towards this end, we appeal to a result in combinatorial optimization~\cite[Theorem 6.5.11]{grotschel2012geometric}. This theorem states that, if we can optimize an objective function over the M-polytope using a separation oracle that runs in polynomial-time, then an algorithmic implementation of Caratheodory's theorem can be performed in polynomial-time in the number of variables. Our result in Theorem~\ref{thm:Oracle} proves that such a polynomial-time separation oracle exists, and hence~\cite[Theorem 6.5.11]{grotschel2012geometric} ensures that the decomposition in~\eqref{eq:carathedory_decomp} can be performed in polynomial-time. 

The remainder of this subsection is devoted to describing the decomposition algorithm~\cite[Theorem 6.5.11]{grotschel2012geometric} and explaining why we can apply it to our M-polytope. For this, we need to explicitly mention some properties of polyhedra.

\begin{defin}
 The dimension of a polyhedron $P \subseteq \mbb{R}^n$, $dim(P)$ is the maximum number of affinely independent points in $P$ minus 1. If $dim(P) = n$, we say that $P$ is {\bf fully-dimensional}. A polyhedron $P$ is said to {\bf bounded} if there exists a ball $B$ in $\mbb{R}^n$ centered around the origin with radius $r < \infty$ such that $P \subseteq B$.
 \end{defin} 

 \begin{defin}
 A polyhedron $P \subseteq \mbb{R}^n$ is called {\bf rational} if all its vertices and at least one point in its interior belong to $\mbb{Q}^n$. A polyhedron $P$ is called {\bf well-described}, if a finite number of bits is needed to encode a single constraint of the polyhedron.
 \end{defin} 

\begin{defin}
 The subset $F$ is called a {\bf face} of polyhedron $P \subseteq \mbb{R}^n$, if there exists an inequality such that $a^T x \leq a_0,\ \forall x \in P$ and $F = \{x \in P | a^T x = a_0\}$. We say that the inequality $a^T x \leq a_0$ defines the face $F$. 
 \end{defin} 

\begin{defin}\label{def:facet}
 A face $F$ of polyhedron $P \subseteq \mbb{R}^n$, is called a {\bf facet} of $P$ if $dim(F) = dim(P) - 1$. If the polyhedron $P$ is fully-dimensional, then each non-redundant inequality constraint of $P$ defines a facet of $P$. 
 \end{defin}

 It is not difficult to see that the M-polytope is rational, well-described and fully-dimensional. To see full-dimensionality, note that the set of matchings (vertices) such that each selects only one edge in the graph, together with the all-zero matching, form a set of affinely independent points with $|E_G| + 1$ elements. As a result (by Definition~\ref{def:facet}) each constraint in~$(1e)-(1g)$ defines a facet of the M-polytope.

To describe the decomposition algorithm for our polytope, we need to shorthand two abstract oracles that we will use with their shorthand (SEP and OPT), extensively.
\begin{enumerate}
\item \underline{$SEP(P, y)$} denotes the separation oracle that takes a polyhedron $P$ and a point $y$. This oracle determines if $y \in  P$ or else returns a constraint of $P$ that is violated by $y$. If $P$ is fully-dimensional, then this returned constraint defines a facet of $P$ as aforementioned above. Note that $P$ might not be explicitly defined to the oracle with a set of constraints, but rather as an object (e.g., a graph) and a condition on the object (e.g., minimum odd cut is greater than some value). An example of such an oracle is the result in Theorem~\ref{thm:Oracle}.

\item \underline{$OPT(P, c)$} denotes the optimization oracle that given a polyhedron $P$ and an affine objective function parameterized with $c$, maximizes $c^Tx$ over $x \in P$. If $P$ is bounded then for any $c$, the oracle returns a vertex of $P$ at which $c^Tx$ is maximized.
\end{enumerate}

\begin{figure}
\centering
\includegraphics[width=0.7\columnwidth]{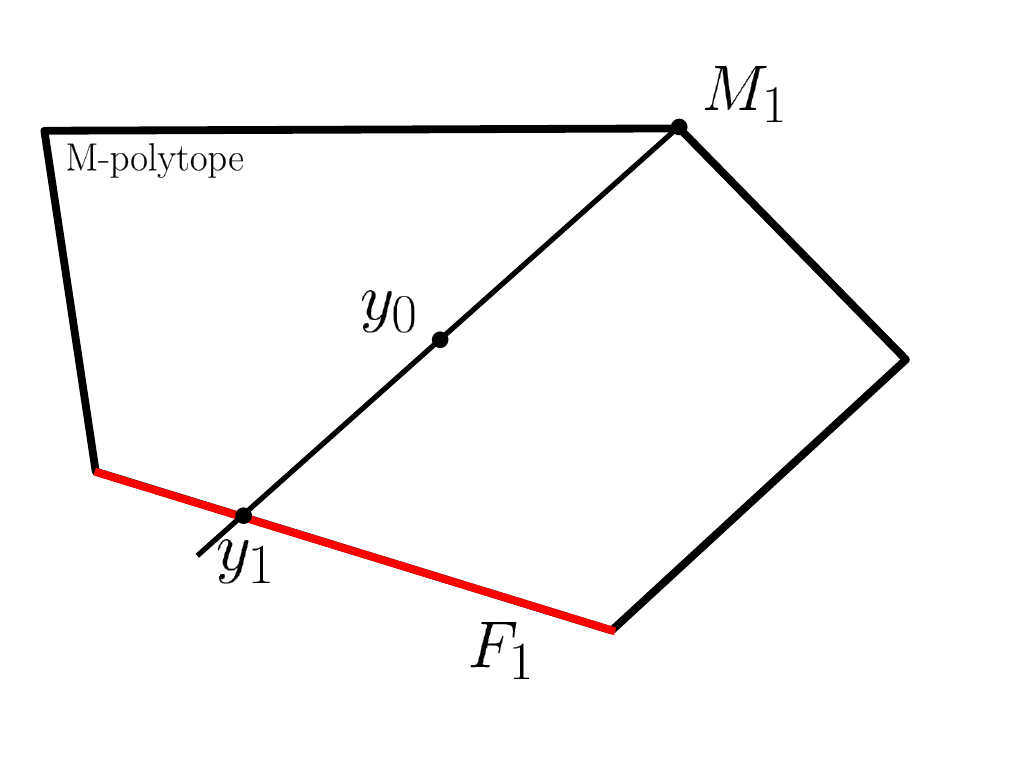}
\caption{An example of the decomposition of $y_0$ into $M_1$ and $y_1$. The red color represents the facet $F_1$ that $y_1$ belongs to.}
\label{fig:polyhedra}
\end{figure}

We are now ready to describe the algorithm used for the decomposition in~\eqref{eq:carathedory_decomp} over the M-polytope. The exact details are described in Algorithm~\ref{algo:PolyCartheodory}.
The skeleton of Algorithm~\ref{algo:PolyCartheodory} is the following:
\begin{enumerate}
\item We start with our desired point to decompose $y_0 = \hat\lambda$ and find any vertex $M_1$ of the M-polytope (using OPT).
\item Maximize along the line connecting $y_0$ and $M_1$ in the M-polytope in the direction $y_0 - M_1$. The maximizer, denoted as $y_1$, is the intersection of the line and some boundary of the M-polytope.  A simple illustration of $y_0$, $y_1$ and $M_1$ is shown in Fig.~\ref{fig:polyhedra}.
\item Since $y_0$ is in between $M_1$ and $y_1$ on a line, then we can write $y_0 = \theta_1 M_1 + (1-\theta_1)y_1$ for $0 \leq \theta \leq 1$. 
\item Next, we find a facet $F_1$ (defined as $a_1^T x = b_1$) of the M-polytope containing $y_1$ by finding the constraint separating $y_1 + \varepsilon (y_0 - M_1)$ from the M-polytope for some $\varepsilon > 0$ (using SEP).
\item We want to find a vertex of this facet so that we maximize the objective function $c_1 = a_1$ (using OPT) to get a vertex $M_2$ in the $F_1$.
\item Since both $y_1$ and $M_2$ belong to $F_1$ then the line projected along them to get $y_2$ also belongs to $F_1$.
\item Repeat Steps 2) to 6) for $i > 1$ until we hit a vertex in the end. Note that in Step 5) to get $M_{i+1}$ that belongs to the intersection of all facets $F_1, F_2, \cdots , F_i$, we use $c_i = \sum_{j=1}^i a_i$ to ensure that the vertex satisfies $a_j^T x = b_j, \ \forall j \in [1:i]$. 
\item At the end, we have that for each $i \in [1:|E_G|+1]$,
\[
	y_{i-1} = \theta_i y_i + (1-\theta_i) M_i,\qquad 0 \leq \theta_i \leq 1.
\]
Thus, by applying recursion, we can express our desired point $y_0$ as
\[
	y_0 = \sum_{i =1}^{|E_G|+1} \varphi_i M_i,
\]
with 
\[
	\varphi_i = \theta_i\prod_{j=1}^{i-1} (1-\theta_j).
\]
\end{enumerate}

\begin{algorithm*}\label{algo:PolyCartheodory}
  \caption{Decomposition into Caratheodory points}
\label{algo:PolyCartheodory}
  \begin{algorithmic}[1]
  	\renewcommand{\algorithmicrequire}{\textbf{Input:}}
  	\renewcommand{\algorithmicensure}{\textbf{Output:}}
  	\Require Graph $G = (V_G, E_G, \hat \lambda)$
	 \Ensure $\{\varphi_i\}$, $\{M_i\}$
  	\State $y_0 = \hat\lambda$
  	\State $P$ = M-polytope of $G$
  	\State $c_0 = 1_{|E_G|\times 1} \quad$   {\blue // Set first objective function to any value (all ones here)}
	\ForEach{$i \in [1:|E_G|+1]$}
	  	\State $M_i = OPT(P,c_{i-1})\qquad \qquad$ {\blue // Get vertex maximizing $c_{i-1}$}
	  	\State {\blue // Maximize in the direction of $y_{i-1} - M_i$ to get the point $y_i$ on the boundary of $P$}
	  	\State $y_i = OPT(P \cap \text{line}(y_{i-1}, M_i), y_{i-1} - M_i)\qquad$ 
	  	\State {\blue // Convex combination of $M_i$ and $y_{i}$ gets $y_{i-1}$}
	  	\State $\theta_i = \frac{M_i - y_{i}}{y_{i-1} - y_{i}}\qquad $ {\blue // Solve for $0 \leq\theta_i \leq 1$ : $\theta_i M_i + (1-\theta_i) y_{i} = y_{i-1}$ }
		\State {\blue // Get the facet containing $y_{i}$. Perturb $y_{i}$ by a small amount in the direction of $y_{i-1} - M_i$ to get a point outside of the polyhedron, and then apply the separation oracle.}
		\State $a_i^T, b_i = SEP(P,y_{i} + \varepsilon \frac{y_{i-1} - M_i}{|y_{i-1} - M_i|})$ {\blue // $a_i^T x = b_i$ defines a facet containing $y_{i}$}
		\State {\blue //We want the new vertex in the next iteration to be in the intersection of all facets $F_j$ visited before. This is done by adding all the inequalities defining these facets}
		\State $c_i = \sum_{j=1}^{i} a_j$
	\EndFor
	\ForEach{$i \in [1:|E_G|+1]$}
	\State	$\varphi_i = \theta_i\prod_{j=1}^{i-1} (1-\theta_j)$
	\EndFor
  \end{algorithmic}
\end{algorithm*}

Note that Algorithm~\ref{algo:PolyCartheodory} above iterates over the oracles for a polynomial number of times equal to $|E_G|+1$, since at each time, we are restricting our search with a new equation ($a_j^T x=  b_j$), i.e., at every iteration we decrease the dimension of the currently considered polytope by 1. Therefore, if the oracles SEP and OPT can run in polynomial time, then the algorithm is polynomial in time. Fortunately, the M-polytope has a polynomial-time separation oracle (by Theorem~\ref{thm:Oracle}) and a polynomial-time optimization oracle (by Theorem~\ref{cor:HDCapPoly}). Thus, by consequence Algorithm~\ref{algo:PolyTimeOr} runs in polynomial-time in the number of relay nodes $N$.

\subsection{Post-processing for directional matchings}
We now need to utilize the matchings $\{M_k\}$ and their activation times $\{\varphi_k\}$ output by the algorithm discussed in the previous subsection to construct network states and find their activation times such that each link $i {\to} j$ is activated for a duration $\lambda_{ji}$ (output by $\rm{P1}$).
We can perform this decomposition in polynomial-time by iterating over the edges of the undirected graph $G$ constructed in the previous subsection.
For notational ease, we can rewrite each of the matchings $M_k$ discovered in the previous subsection as a lower triangular matrix $\widetilde{M}_k \in \mathbb{R}^{(N+2) \times (N+2)}$, where
\[
	\widetilde{M}_k(j,i) = 
	\begin{cases}
		 M_k(e) & (i-1,j-1) \in E_G,\ i < j \\ 
		 0 & \text{otherwise},
		\end{cases}	
\]
where $i-1$ and $j-1$ are used since the matrix entries are matched with positive numbers while our nodes are indexed from $0$.
The rows of $\widetilde{M}_k$ represent the receiving modules of the nodes, while the column indexes represent the transmitting modules of the nodes.
As an illustrative example, consider a Gaussian HD 1-2-1 network with $N = 2$ relays, where the undirected graph $G$ has edge set $E_G = \{(0,1),(0,2), (1,2), (1,3), (2,3)\}$. 
For the matching $M_k$ shown below, we have the corresponding $\widetilde M_k$
\begin{align*}
 M_k = 
    \begin{bmatrix}
      1 \\ 0 \\ 0 \\ 0 \\ 1
    \end{bmatrix} 
    \quad \implies \quad
    \widetilde M_k = 
    \begin{bmatrix}
     0 & 0 & 0 & 0 \\
     1 & 0 & 0 & 0 \\ 
     0 & 0 & 0 & 0 \\ 
     0 & 0 & 1 & 0
    \end{bmatrix}.
\end{align*}	
\begin{algorithm*}
  \caption{Constructing digraph matchings from undirected graph matchings}
\label{algo:directions}
  \begin{algorithmic}[1]
  	\renewcommand{\algorithmicrequire}{\textbf{Input:}}
  	\renewcommand{\algorithmicensure}{\textbf{Output:}}
  	\Require Graph $G = (V_G, E_G, \hat \lambda),\ \{\widetilde M_k\}_{k=1}^{|E_G|+1},\ \{\varphi\}_{k=1}^{|E_G|+1}$
	 \Ensure $\{\varphi'_k\}$, $\{\widetilde M'_k\}$
  	\State $K = |E_G| + 1$
  	\State $\widetilde M'_k = \widetilde M_k,\ \ \varphi'_k = \varphi_k$, $\qquad \forall k \in [1:|E_G| + 1]$
	\ForEach{$(i,j) \in E_G$}
		\State $\omega = 0$
		\ForEach{$k \in [1:K]$}
			\If{$\widetilde M'_k(j,i) == 1$}
				\If {$ \omega + \varphi'_k < \lambda_{ji}$} \quad {\blue // All matchings that contain $(i,j)$ connection so far are less than $\lambda_{ji}$ }
					\State $\omega = \omega + \varphi'_k$
				\ElsIf {$\omega + \varphi'_k = \lambda_{ji}$} \quad {\blue // All matchings that contain $(i,j)$ connection so far are exactly $\lambda_{ji}$ }
					\State  {\blue // For all remaining matchings containing $(i,j)$ assign to $j \to i$}
					\ForEach{$\ell \in [k+1:K]\ s.t.\ \widetilde M'_\ell(j,i) == 1$}
						\State $\widetilde M'_\ell(j,i) = 0,\quad \widetilde M'_\ell(i,j) = 1$
					\EndFor
					\State $\mathbf{Break}$
				\Else
					\State  {\blue // Make another copy ($K+1$) of the current state, and assign $j \to i$ instead of $i \to j$} 
					\State $\widetilde M'_{K+1} = \widetilde M'_k$ {\blue // Make a new copy at the end of the list}				
					\State $\varphi'_{K+1} = \omega + \varphi'_{k} - \lambda_{ji}$, \qquad $\varphi'_{k} = \lambda_{ji} - \omega$
					\State $\widetilde M'_{K+1}(j,i) = 0,\quad \widetilde M'_{K+1}(i,j) = 1$
					\State  {\blue // For all remaining matchings in $[k+1:K]$containing $(i,j)$ assign to $j \to i$}
					\ForEach{$\ell \in [k+1:K]\ s.t.\ \widetilde M'_\ell(j,i) == 1$}
						\State $\widetilde M'_\ell(j,i) = 0,\quad \widetilde M'_\ell(i,j) = 1$
					\EndFor
					\State $K = K + 1$ \quad {\blue // Increase the number of states by one due to the copy created in line 16}
					\State $\mathbf{Break}$
				\EndIf
			\EndIf
		\EndFor
	\EndFor
  \end{algorithmic}
\end{algorithm*}
With this notation, we can design an algorithm that generates the $\{\lambda_s\}$ optimal for~\eqref{eq:apprCap} such that they collectively activate each link $i \to j$ for the duration $\lambda_{ji}$ output by the LP $\rm{P1}$. The formal procedure of the algorithm is described in Algorithm~\ref{algo:directions}.
The main idea behind Algorithm~\ref{algo:directions} is the following. From~\eqref{eq:carathedory_decomp}, the definition of $\widetilde M_k$ and the LP $\rm{P1}$, we know that $\forall (i,j) \in [0:N{+}2],$ such that $i < j$, we have
\[
\sum_{k=1}^{|E_G| + 1} \varphi_k\ \widetilde M_k(j,i) = \hat \lambda_{ij} \stackrel{(\text{P1})}= \lambda_{ji} + \lambda_{ij}. 
\]
Thus, for each connection $(i,j)$ in the network, we just need to break the matchings $\{\widetilde M_k\}$ into two sets contributing to the activation of the links $i \to j$ and $j \to i$.
Before processing a connection $(i,j)$, the default direction is $i \to j$.
The algorithm iterates over each connection $(i,j)$, and adds up the activation times for the matchings one by one until the sum exceeds $\lambda_{ji}$. For the remaining matchings, we change the assigned direction to $j \to i$. The matching $\widetilde{M}_k(j,i)$ that caused the sum to exceed $\lambda_{ji}$ is split into two copies, one where the direction is $i \to j$ and the other with $j \to i$.

Note that, in each iteration over the elements in $E_G$, we split at most one matching (state). Thus, starting with $|E_G|+1$ matchings, we end up with at most $2|E_G| + 1$ matchings. 
Moreover, the inner loop iterates over at most $2|E_G|+1$ matchings. Thus, the algorithm runs in $O(|E_G|^2)$ time which in the worst case is $O(N^4)$ for a network with $N$ relays.


\appendices

\section{Proof of Theorem~\ref{thm:ThLinkAct}}
\label{sec:ProofTh1}
We start this section by taking note of an observation in~\cite{EzzeldinISIT2018}, which states that,
 for a fixed $\lambda_s$, the inner minimization in $\msf{C}_{\rm cs,iid}$ in~\eqref{eq:apprCap} is the standard min-cut problem over a graph with link capacities given by 
\begin{align}
\ell_{j,i}^{(s)} =\left( \sum_{\substack{ s:\\ j \in s_{i,t},\ i \in s_{j,r} }} \lambda_s \right) \ell_{j,i}.
\label{eq:NewLinkCap}
\end{align}
We can therefore leverage the fact that the dual of the min-cut problem is the max-flow problem, and use strong duality to replace the inner minimization in~\eqref{eq:apprCap} with the max-flow problem over the graph with link capacities defined in~\eqref{eq:NewLinkCap}. 
With this, we obtain
\begin{align*}
\rm{P0}&{\rm\ :} \ \msf{C}_{\rm cs,iid} = \max \sum_{j =1}^{N+1} F_{j,0}& \nonumber\\
& (0a)\quad 0 \leq F_{j,i} \leq  \ell_{j,i}^{(s)}, \quad \forall (i,j) \in [0{:}N]\times[1{:}N{+}1], \\
&(0b) \sum_{j\in[1:N{+}1]\backslash\{i\}} F_{j,i} = \sum_{k\in[0:N]\backslash\{i\}} F_{i,k}, \quad \forall i \in [1:N],\\
& (0c)\quad \sum_{s} \lambda_s \leq 1, \\
& (0d)\quad\ \ \lambda_s \geq 0,
\end{align*}
where $F_{j,i}$ represents the flow from node $i$ to node $j$, and where the constraint in $(0b)$ ensures that flow is conserved at each relay node.
Thus, to prove Theorem~\ref{thm:ThLinkAct}, we need to show that the linear program $\rm{P1}$ is equivalent to the linear program $\rm{P0}$.

We start by noting that all variables, except $\lambda_s$ and $\lambda_{ij}$, are the same in the two linear programs $\rm{P0}$ and $\rm{P1}$. Thus, to make $\rm{P0}$ and $\rm{P1}$ equivalent, we want to find a bijective mapping between $\lambda_s$'s and $\lambda_{ij}$'s.

If we observe a state $s$ in a Gaussian HD 1-2-1 network and hence in $\rm{P0}$, we see that a state does not activate two adjacent links, i.e., a state represents a matching\footnote{A matching in a graph is a set of graph edges such that each vertex (node) in the graph has at most one edge from the set connected to it.} in the graph representing the network topology.
Furthermore, the constraints in $(0a)$, $(0c)$ and $(0d)$ in $\rm{P0}$ suggest that the link activation times should be in the convex hull of the 0-1 activations representing a matching.
To state this more formally, let $F$ be a matrix that is populated by the flow variables $F_{j,i}$ in $\rm{P0}$, and let $L$ be a matrix populated by the link capacities $\ell_{j,i}$ in $\rm{P0}$.
For each state $s$, let $M_s$ be a binary matrix such that its $(j,i)$-th component is $1$ if the link from node $i$ to node $j$ is activated by state $s$, and $0$ otherwise.
Then, we can write the constraints in $(0a)$ and $(0c)$ in $\rm{P0}$ as 
\begin{subequations}
\label{eq:const_P0}
\begin{align}
& (0a): \quad F \leq \left(\sum_{s}\lambda_s M_s \right) \odot L, \label{eq:const_P0a}
\\& (0c): \quad \sum_{s} \lambda_s = 1, \label{eq:const_P0b}
\end{align}
\end{subequations}
where the operator $\odot$ denotes the element-wise multiplication (Hadamard product).
Note that we made the sum of the state probabilities in~\eqref{eq:const_P0b} equal to $1$, and this is without loss of generality. 
This is because one valid $M_s$ is populated by all zeros, and hence we can associate to this all-zero matrix a weight so that the sum equals $1$, without changing the main inequality in~\eqref{eq:const_P0a}.
Now, if we consider $\rm{P1}$, then the constraint in $(1a)$ can be written as 
\begin{align}\label{eq:const_P1}
	F \leq \Lambda \odot L,
\end{align}
where $\Lambda$ is populated by the variables $\lambda_{ji}$.

From \eqref{eq:const_P0} and \eqref{eq:const_P1}, we can see that the constraint needed in $\rm{P1}$ is to have $\Lambda$ in the convex hull of $\{M_s\}$,
i.e., we want the link activations (or weights) that are in the convex hull of points representing matchings in a graphs.
A result~\cite{edmonds1965} by Edmonds for undirected graphs characterizes the constraints that represent the Matching polytope (M-polytope). The M-polytope of an undirected graph $G$ is the polytope that has all matchings as its extreme points.
By massaging this result to apply to the directed graph in our problem, we define $\hat{\lambda}_{ij} = \lambda_{ij} + \lambda_{ji}$ for all $i < j$ and we apply Edmonds's constraints on $\hat{\lambda}_{ij}$.
With this, we precisely get the linear program in $\rm{P1}$. In particular, the M-polytope characterized by Edmonds defines the constraints in $(1e)$ and $(1f)$ on $\hat{\lambda}_{ij}$. The mapping from $\hat{\lambda}$ to $\lambda_{ij}$ is defined by the constraint in $(1c)$ in $\rm{P1}$. 
This concludes the proof of Theorem~\ref{thm:ThLinkAct}.

\section{Equivalence between PM-polytopes in~\eqref{eq:PMP_} and \eqref{eq:PMP}}\label{sec:PM_equiv}

We note that in~\eqref{eq:PMP_} and \eqref{eq:PMP}, the first two constraints are the same. 
Thus, what we need to show is the equivalence between the third constraints in~\eqref{eq:PMP_} and \eqref{eq:PMP}.
In other words, given an undirected graph $G = (V_G,E_G,\hlambda)$, with 
\begin{align}\label{eq:const_12_PMP}
\begin{array}{rll}
 \hlambda(e) &\geq 0,& \forall e \in E_G,\\
   	\hlambda(\delta_G(v)) &= 1,& \forall v \in V_G,
   	\end{array}
 \end{align} 
 we want to show that for $\forall S \subseteq V_G, |S|\ \text{odd}$, we have that
 \begin{align*}
    \hlambda(E_G(S)) \leq \dfrac{|S|-1}{2} \iff \hlambda(\delta_G(S)) \geq 1.
 \end{align*}
A key property used to prove both directions is the fact that for any $S \subseteq V_G$, we have -- by simple counting -- that
\begin{equation}\label{eq:edge_div}
	\sum_{v \in S} \hlambda(\delta_G(v)) = \hlambda(\delta_G(S)) + 2 \hlambda(E_G(S)).
\end{equation}
\underline{$\Longrightarrow$ direction:}
Let $S$ be an arbitrary subset of $V_G$ with odd cardinality. Then, we have that
\begin{align*}
  |S| &\stackrel{\eqref{eq:const_12_PMP}}= \sum_{v\in S} \hlambda(\delta_G(v)) \\
  	 	&\stackrel{\eqref{eq:edge_div}}= \hlambda(\delta_G(S)) + 2 \hlambda(E_G(S)) \\
  	 	&\leq \hlambda(\delta_G(S)) + 2 \frac{|S|-1}{2}.
\end{align*}
The above expression can be rewritten to give that $\hlambda(\delta_G(S)) \geq 1$, which is precisely what we intended to show.


\noindent \underline{$\Longleftarrow$ direction:}
Let $S$ be an arbitrary subset of $V_G$ with odd cardinality. Then, we have that
\begin{align*}
  |S| &\stackrel{\eqref{eq:const_12_PMP}}= \sum_{v\in S} \hlambda(\delta_G(v)) \\
  	 	&\stackrel{\eqref{eq:edge_div}}= \hlambda(\delta_G(S)) + 2 \hlambda(E_G(S)) \\
  	 	&\geq 1 + 2 \hlambda(E_G(S)),
\end{align*}
which, when reorganized, gives that $\hlambda(E_G(S)) \leq (|S|-1)/2$. This concludes the proof that~\eqref{eq:PMP_} and~\eqref{eq:PMP} are equivalent.

\section{Proof that $\widetilde\lambda$ in~\eqref{eq:lambdaTilde} satisfies~\eqref{eq:perfect_graph}}
\label{app:RelPMandM}

We want to show that
 given $\hlambda$ in the M-polytope of $G$, then $\widetilde\lambda$ constructed as in~\eqref{eq:lambdaTilde} satisfies~\eqref{eq:perfect_graph} -- and hence it is in the PM-polytope of $\widetilde G$.

For any $S \subseteq V_{\widetilde G}$ we define $S_a = S \cap V_G$ and $S_b' = S \cap V_{G'}$. 
We let $S_a'$ be the copy of $S_a$ in $V_{G'}$ and similarly, we let $S_b$ be the copy of $S_b'$ in $V_G$. Note that, $S = S_a \cup S'_b$. 
Fig.~\ref{fig:Gtilde} provides an illustration of these sets.
We now consider two different cases for $S$ with odd cardinality and, for each of them, prove that the constraint in~\eqref{eq:perfect_graph} is indeed satisfied.
This will show that the construction of $\widetilde\lambda$ in~\eqref{eq:lambdaTilde} satisfies the constraint in~\eqref{eq:perfect_graph}.
\begin{enumerate}
\item {\bf Case 1: $S_b'  = \emptyset$}. In this case, since $S = S_a \cup S'_b$, then $|S| = |S_a|$ is odd, and we have that
	\begin{align*}
	   \widetilde{\lambda}(\delta_{\widetilde G}(S)) &= \widetilde{\lambda}(\delta_{\widetilde G}(S_a)) \\
													 & \stackrel{(a)}= \sum_{v \in S_a} \widetilde{\lambda}(\delta_{\widetilde G}(v)) - 2 \widetilde{\lambda}(E_{\widetilde G}(S_a)) \\
										     		 &\stackrel{(b)}= |S_a| - 2 \widetilde{\lambda}(E_{\widetilde G}(S_a)) \\
	   												 &\stackrel{(c)}\geq |S_a| - 2 \frac{|S_a| - 1}{2} = 1,
	\end{align*}
where: (i) the equality in $(a)$ follows by using counting arguments (see also~\eqref{eq:edge_div} in Appendix~\ref{sec:PM_equiv});
(ii) the equality in $(b)$ follows from the fact that our construction of $\widetilde \lambda$ satisfies the second constraint in~\eqref{eq:PMP};
(iii) the inequality in $(c)$ follows from the fact that, by our construction in~\eqref{eq:lambdaTilde} $\widetilde{\lambda}(E_{\widetilde G}(S_a)) = \hat{\lambda}(E_{G}(S_a))$ and for any $S_a \subseteq V_G$, from the third constraint in~\eqref{eq:MP}, we have that $\hlambda(E_G(S_a))\leq \dfrac{|S_a|-1}{2}$. 
Thus, for this case the constraint in~\eqref{eq:perfect_graph} is satisfied. 

\begin{figure}
\centering
\includegraphics[width=\columnwidth]{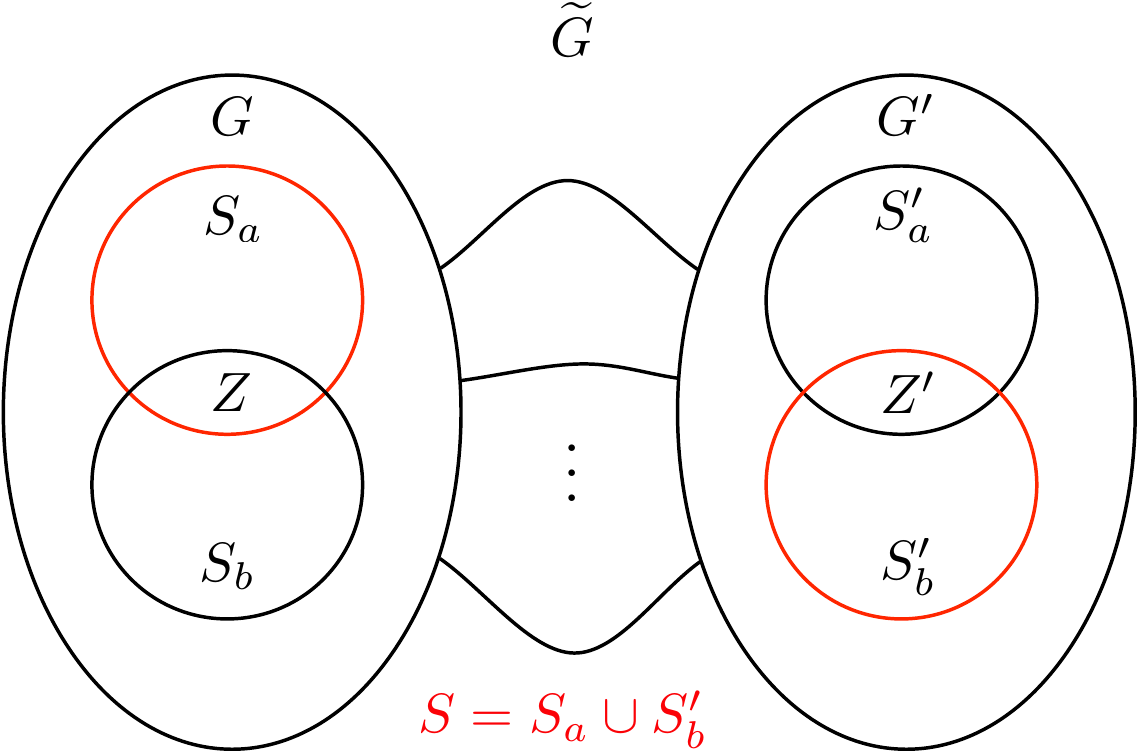}
\caption{Illustration for the different sets used in Appendix~\ref{app:RelPMandM}.}
\label{fig:Gtilde}
\end{figure}
\item  {\bf Case 2: $S_b' \neq \emptyset$}. We start by noting that (see also Fig.~\ref{fig:Gtilde})
	\begin{align*}
		|S| &= |S_a| + |S_b'| \\
			&= |S_a \backslash S_b| + |\underbrace{S_a \cap S_b}_{Z}|+ |S_b' \backslash S_a'| + |\underbrace{S_a' \cap S_b'}_{Z'}| \\
			&= |S_a \backslash S_b| + |S_b' \backslash S_a'| + 2|S_a \cap S_b|.
	\end{align*}
	Thus, since $|S|$ is odd, then one (and only one) among $|S_a \backslash S_b|$ and $|S_b' \backslash S_a'|$ is odd. 
	Without loss of generality, assume that $|S_a \backslash S_b|$ is odd. Then, we have that (see also Fig.~\ref{fig:Gtilde}) 
\begin{align*}
\widetilde{\lambda}(\delta_{\widetilde G}(S)) &= \widetilde{\lambda}(\delta_{\widetilde G}(S_a\backslash S_b)) 
+ \widetilde{\lambda}(\delta_{\widetilde G}(Z))\\
		& \quad - 2 \widetilde{\lambda}(E_{\widetilde G}(S_a\backslash S_b, Z )) - \widetilde{\lambda}(E_{\widetilde G}(Z, Z' ))\\
& \quad + \widetilde{\lambda}(\delta_{\widetilde G}(S_b'\backslash S_a')) + \widetilde{\lambda}(\delta_{\widetilde G}(Z'))\\
&  \quad - 2 \widetilde{\lambda}(E_{\widetilde G}(S_b'\backslash S_a', Z' )) - \widetilde{\lambda}(E_{\widetilde G}(Z, Z' ))\\
		&\stackrel{(a)}= \widetilde{\lambda}(\delta_{\widetilde G}(S_a\backslash S_b)) + \widetilde{\lambda}(\delta_{\widetilde G}(S_b'\backslash S_a')) \\
		& \quad + \widetilde{\lambda}(\delta_{\widetilde G}(Z)) - \beta^\star\\
		& \quad + \widetilde{\lambda}(\delta_{\widetilde G}(Z')) - \beta^\star  \\
		& \stackrel{(b)}= \widetilde{\lambda}(\delta_{\widetilde G}(S_a\backslash S_b)) + \widetilde{\lambda}(\delta_{\widetilde G}(S_b'\backslash S_a')) \\
		&\quad + 2\widetilde{\lambda}(\delta_{\widetilde G}(Z)) - 2\gamma^\star \\
%
		&\geq \widetilde{\lambda}(\delta_{\widetilde G}(S_a\backslash S_b)) + \widetilde{\lambda}(\delta_{\widetilde G}(S_b'\backslash S_a')),
	\end{align*}
where: (i) the equality in $(a)$ follows by defining
\begin{align*}
\beta^\star =& \left[\widetilde{\lambda}(E_{\widetilde G}(S_b'\backslash S_a', Z' )) + \widetilde{\lambda}(E_{\widetilde G}(Z, Z' )) \right .
\\ & \left .+ \widetilde{\lambda}\left(E_{\widetilde G}(S_a\backslash S_b, Z )\right)  \right];
\end{align*}
(ii) the equality in $(b)$ follows by noting that $G'$ is a copy of $G$ and hence $\widetilde{\lambda}(\delta_{\widetilde G}(Z))=\widetilde{\lambda}(\delta_{\widetilde G}(Z'))$, $\widetilde{\lambda}\left(E_{\widetilde G}(S_a'\backslash S_b', Z' )\right) = \widetilde{\lambda}\left(E_{\widetilde G}(S_a\backslash S_b, Z )\right)$ and similarly, $\widetilde{\lambda}\left(E_{\widetilde G}(S_b'\backslash S_a', Z' )\right) = \widetilde{\lambda}\left(E_{\widetilde G}(S_b\backslash S_a, Z )\right)$; thus, with this, we have
\begin{align*}
\beta^\star = \gamma^\star =& \left[\widetilde{\lambda}(E_{\widetilde G}(S_a\backslash S_b, Z )) + \widetilde{\lambda}(E_{\widetilde G}(Z, Z' )) \right.
\\ & \left.+ \widetilde{\lambda}\left(E_{\widetilde G}(S_b\backslash S_a, Z )\right)\right].
\end{align*}
From the series of inequalities above, we have that
	\begin{align}\label{eq:S_a_b}
		\widetilde{\lambda}(\delta_{\widetilde G}(S)) \geq \widetilde{\lambda}(\delta_{\widetilde G}(S_a\backslash S_b)) + \widetilde{\lambda}(\delta_{\widetilde G}(S_b'\backslash S_a')).	
	\end{align}

Since we assumed, without loss of generality, that $|S_a\backslash S_b|$ is odd and by using similar arguments as in Case 1 (since $S_a\backslash S_b \subset V_G$), then it follows that
\begin{align*}
& \widetilde{\lambda}(\delta_{\widetilde G}(S_a\backslash S_b)) \geq 1 
\\& \implies \widetilde{\lambda}(\delta_{\widetilde G}(S)) \geq 1.
\end{align*}
Thus, for this case the constraint in~\eqref{eq:perfect_graph} is satisfied. 
\end{enumerate}
The above analysis shows that, given $\hlambda$ in the M-polytope of $G$, then $\widetilde\lambda$, with the construction in~\eqref{eq:lambdaTilde}, satisfies~\eqref{eq:perfect_graph} -- and hence it is in the PM-polytope of $\widetilde G$.

\section{Proof of Theorem~\ref{thm:odd_cuts}}\label{sec:odd_cuts_thm}
We here prove Theorem~\ref{thm:odd_cuts} -- which is a simplified version of the result in~\cite{padberg1982odd} -- by following the logic in~\cite[Theorem~25.6]{schrijver2003combinatorial}.
In particular, the goal of the proof is to show that, for {\em any} odd cut $U$ in the graph $\widetilde G$, we can always consider any Gomory-Hu tree $T$ of $\widetilde G$, and find an odd cut $W_{f^\star}$ inside it, such that $\delta_{\widetilde G}(U) \geq \delta_{\widetilde G}(W_{f^\star})$. Since this is true for all odd cuts, then it is also true for the minimum odd cut. 
This implies that one of the odd cuts in a Gomory-Hu tree is actually a minimum odd cut in $\widetilde G$. In the remainder of this appendix, we formalize this argument.

Let $T = (V_{\widetilde G}, F, \alpha_T)$ be a Gomory-Hu tree of the graph $\widetilde G = (V_{\widetilde G}, E_{\widetilde G}, \widetilde \lambda)$, where we assume that $|V_{\widetilde G}|$ is even.
Then, $\forall f \in F$, we let $W_f$ be one of the two components of $T\backslash f$.
Let $U$ be an arbitrary odd cut in $\widetilde G$. Applying $U$ to the Gomory-Hu tree $T$ gives us $\delta_T(U)$, which we refer to as the cut-set of $U$, i.e., the set of edges that separate $U$ from $U^c$ in the Gomory-Hu tree $T$.
Note that $\delta_T(W_f) = f$. Thus, we have that
\begin{align}\label{eq:cut_prop_symdiff}
    \delta_T(U) &= \symdiff_{f \in \delta_T(U)} \{f\}  \nonumber
\\&= \symdiff_{f \in \delta_T(U)} \delta_T(W_f)  \nonumber
\\ & \stackrel{(a)}= \delta_T\left(\symdiff_{f \in \delta_T(U)} W_f\right),
\end{align}
where $A \Delta B$ is the symmetric difference between the sets $A$ and $B$, and where in $(a)$ we appeal to the property of cuts that for any two sets $S_1, S_2 \subset V_{\widetilde G}$, we have that
\[
\delta(S_1) \Delta \delta(S_2) = \delta\left(S_1 \Delta S_2 \right).
\] 
Since cuts in a tree have unique cut-sets (excluding complements), then from~\eqref{eq:cut_prop_symdiff}, we have that 
\[
\displaystyle\symdiff_{f \in \delta_T(U)} W_f = U\ \ \text{ or }\ \ \displaystyle\symdiff_{f \in \delta_T(U)} W_f = U^c.
\]
Moreover, since $|V_{\widetilde G}|$ is even and, by assumption, $|U|$ is odd, then $|U^c|$ is also odd. Thus, in any case, we have that $\left|\displaystyle\symdiff_{f \in \delta_T(U)} W_f\right|$ is odd.
Finally, the last step in the proof is to appeal to the following property that relates the cardinality of two sets to their symmetric difference. For any two sets $A$ and $B$, we have that
\begin{align}
	|A \Delta B| = |A| + |B| - 2|A \cap B|.
\end{align} 
Thus, if $|A \Delta B|$ is odd, then either $|A|$ or $|B|$ is odd. This extends to the case, when we have $\Delta_{i=1}^M A_i$. In this case, at least one of the sets $A_i$ has an odd cardinality. Applying this observation to the set $\displaystyle\symdiff_{f \in \delta_T(U)} W_f$, we have that for some $f^\star \in \delta_T(U)$, $|W_{f^\star}|$ is odd.

Let $f^\star = (s,t)$ and recall that we assumed in our proof that the set $U$ is an arbitrary odd cut in $\widetilde G$. 
Moreover,
the edge $f^\star$ belongs to both $\delta_T(W_{f^\star})$ and $\delta_T(U)$. Thus, both $U$ and $W_{f^\star}$ represent an $s-t$ cut in $\widetilde G$. From the definition of the Gomory-Hu tree (see Definition~\ref{def:GHTree}), $W_{f^\star}$ is a minimum cut for $s-t$. Thus, we have that 
\[	
	\widetilde \lambda(\delta_{\widetilde G}(U)) \geq \widetilde \lambda(\delta_{\widetilde G}(W_{f^\star})).
\]
Furthermore, since $|W_{f^\star}|$ is odd, then we have proved that, for any odd cut $U$ in $\widetilde G$, we can find one cut $W_{f^\star}$ in $T$ (cuts by removing one edge) that is odd and such that $\widetilde \lambda(\delta_{\widetilde G}(U)) \geq \widetilde \lambda(\delta_{\widetilde G}(W_{f^\star}))$. 
This concludes the proof of Theorem~\ref{thm:odd_cuts}.


\bibliographystyle{IEEEtran}
\bibliography{mmWaveNetwork}

\end{document}